\documentclass{ws-procs9x6}
\usepackage{amsmath,amssymb,natbib}
\usepackage{theorem}

\def\ket#1{{\bigl|#1\bigr\rangle}}
\def\bra#1{{\bigl\langle#1\bigr|}}
\def\braket#1#2{{\bigl\langle#1\bigr|#2\bigr\rangle}}
\def\meas{{\mathrm{meas}}}
\def\law{{\mathrm{law}}}
\def\E{{\mathrm{E}}}
\def\Euro{{\hbox{\it C\hskip-0.75em{\raisebox{0.4ex}{$\scriptstyle =\,$\hskip-0.5em$\scriptstyle =\,$}}}}}
\def\onebb{{\rm  1\mkern-5.4mu I}}

\newtheorem{myconjecture}{Conjecture}
\newtheorem{myassumption}{Assumption}

\begin{document}

\title{On an Argument of David Deutsch}

\author{Richard D. Gill}

\address{Mathematical Institute, University of Utrecht, Netherlands, \&\\
EURANDOM, Eindhoven, Netherlands\\
gill@math.uu.nl\\
http://www.math.uu.nl/people/gill
}

\maketitle

\abstracts{
We analyse an argument of Deutsch, which purports to show 
that the deterministic part of classical quantum theory together with 
deterministic axioms of classical decision theory, together imply that a 
rational decision maker behaves as if the probabilistic part of 
quantum theory (Born's law) is true. We uncover two missing
assumptions in the argument, and show that the argument 
also works for an instrumentalist who is prepared to accept that the 
outcome of a quantum measurement is random in the frequentist sense:
Born's law is a consequence of functional and unitary invariance 
principles belonging to the deterministic part of quantum mechanics.
Unfortunately, it turns out that after the necessary corrections
we have done no more than give an easier proof of Gleason's theorem
under stronger assumptions. However, for some special cases the proof 
method gives positive results while using \emph{different} 
assumptions to Gleason. This leads to the conjecture that the proof 
could be improved to give the same conclusion as Gleason under 
unitary invariance together with a much
\emph{weaker} functional invariance condition.}

\section{Introduction: are quantum probabilities fixed by quantum 
determinism?}

Quantum mechanics has two components: a deterministic component, concerned
with the time evolution of an isolated quantum system; and a stochastic
component, concerned with the random jump which the state of that system
makes when it comes into interaction with the outside world, sending at the
same time a piece of random information into the outside world.  The 
perceived conflict between these two behaviours is `the measurement
problem' as exemplified by Schr\"odinger's cat.

Here we do not resolve this problem but just address the 
peaceful coexistence, or possibly even the harmony,
between the two behaviours.  We will show that  some classical
deterministic quantum mechanical assumptions, together with the
assumption that the outcome of measuring an observable is random, 
uniquely determines the probability distribution of the outcome---harmony indeed. 
More specifically, two generally accepted invariance properties of observables and
quantum systems determine the shape of the probability distribution of
measured values of an observable---namely, the shape specified by 
Born's law.  The invariance properties are connected
to unitary evolution of a quantum system, and to functional transformation
of an observable, respectively.

This work was inspired by Deutsch (1999). There it is claimed
that a still smaller kernel of deterministic classical quantum theory
together with a small part of deterministic decision theory together force a
rational decision maker to behave as if the probabilistic predictions of
quantum theory are true.  In our opinion there are three problems 
with the paper.  The
first is methodological: we do not accept that the behaviour of a rational
decision maker should play a role in modelling physical systems.  We
are on the other hand happy to accept a stochastic component (with a
frequentist interpretation) in physics, so we translate Deutsch's 
axioms
and conclusions about the behaviour of a rational decision maker into
axioms and conclusions about the relative frequency with which various
outcomes of a physical experiment take place.  The second problem is 
that 
it appears that Deutsch has implicitly made use of a further axiom 
of unitary invariance alongside his truely minimalistic collection, 
and needs to greatly strengthen one of the existing assumptions 
concerning functional invariance, from one-to-one functions also to 
many-to-one functions.  Neither addition nor strengthening is 
controversial from a classical deterministic quantum physics point of 
view, 
but both are very substantial from a mathematical point of view. 
The third problem is that the strengthening of the functional 
invariance 
assumption puts us in the 
position that we have assumed enough to apply Gleason's (1957) 
theorem. 
Thus at best, Deutsch's proof is an easy proof of Gleason's theorem 
using an 
extra, heavy, assumption of unitary invariance. 

The fact that 
Deutsch's 
proof is incomplete has been observed 
by Barnum \textit{et al}.\  (2000). However these authors 
did not attempt to reconstruct a correct proof. In the concluding 
section we relate our work to theirs. Wallace (2002, 2003a,b) 
has also studied Deutsch's claims at great length and from
a rather philosophical point of view. 
I did not attempt to relate his work to mine.
The same goes for Saunders (2002).

The paper is organised as follows. In Section \ref{s:ass} we put 
forward functional and unitary invariance assumptions, which are 
usually considered \emph{consequences} of traditional quantum 
mechanics, 
but are here to be taken as \emph{axioms} from which some of the 
traditional ingredients are to be derived, turning the tables so to 
speak. One would like to make the axioms as modest as possible, while 
still obtaining the same conclusions. Hence it is important to 
distinguish between different variants of the assumptions. In 
particular, we distinguish between (stronger) assumptions about the 
complete probability law of outcomes of measurements of observables, 
and (weaker) assumptions about the mean values of those probability 
laws. 
An invariance assumption concerning a class of functions, is weaker, 
if it only demands invariance for a smaller class of functions, and 
in 
particular we distinguish between invariance for all functions, 
including 
many-to-one functions, and invariance just for one-to-one functions.

In Section \ref{s:proof} we prove the required result, Born's law, 
for a special state (equal weight superposition of two eigenstates). 
This case is the central part of Deutsch (1999), who only sketches 
the generalization to arbitrary states. And already, it seems an 
impressive result. We prove the result, \emph{for this special state},
in two forms---in law, and in mean value---the former being stronger 
of course; using appropriate 
variants of our assumptions. Deutsch's proof is incomplete, since he 
only appeals to unitary invariance, while it is clear that a 
functional invariance assumption is also required.

The strengthening of the functional invariance assumption 
can also be used to derive probabilities as well as mean values, 
and it is moreover useful from Deutsch's point of view of rational 
behaviour, 
if one wants to extend in a very natural way the class of games being 
played. Roughly speaking, we extend from the game of buying
a lottery ticket to a game at the roulette table. In the former game
the only question is, how much is one ticket worth. In the latter game
one may make different kinds of bets, and the question is how
much is any bet worth.

However, so far we have only been concerned with a rather special state:
an equal weight superposition of two eigenstates. As mentioned before
Deutsch only sketches the extension to the general case of an
arbitrary, possibly mixed, state.
He outlined a step-by-step argument of successive generalizations.
In Section \ref{s:rest} we follow the same sequence of steps,
strengthening the assumptions as seems to be needed. 

In Section \ref{s:discuss}, we look back at the various versions of our 
assumptions, in the light of what can be got from them.  
We also evaluate the overall result of completing Deutsch's programme.
From a mathematical point of view, it turns out that
we have done no more, at the end of the day, than 
derive the same conclusion as that of Gleason's theorem, while making
stronger assumptions. The payoff has just been a much easier proof. Gleason's
theorem only assumes functional invariance, we have assumed unitary
invariance too. We argue that unitary invariance corresponds to a natural
physical intuition, while functional invariance is something which one could 
not have expected in advance. It \emph{is} supported by experiment,
and \emph{is} theoretically supported in special cases 
(measurements of components of product systems) by locality. 

We conclude with the conjecture that unitary invariance together with
a weakened functional invariance assumption
is sufficient to obtain the same conclusion.

\section{Assumptions: degeneracy, functional invariance, unitary 
invariance}\label{s:ass}

\setcounter{myassumption}{-1}
Recall that a quantum system in a pure state is described or 
represented by a unit 
vector $\ket \psi$ in a Hilbert space, supposed to be 
infinite-dimensional, and that an observable or 
physical quantity is described or represented by a self-adjoint 
(perhaps unbounded) operator $X$ on that space. I shall 
assume that $X$ has a discrete and nondegenerate spectrum; thus there 
is a countably infinite collection of real eigenvalues $x$ and 
eigenstates 
$\ket {X\!=\!x}$, so that one can write $X=\sum_{x} x \ket 
{X\!=\!x}\bra {X\!=\!x}$,
while $\ket\psi=\sum\lambda_{x}\ket {X\!=\!x}$ where 
$\lambda_{x}=\braket{ X\!=\!x}{\psi}$.
Throughout the paper we make the following background assumption:
\begin{myassumption}\textit{Random outcome, in spectrum.}
The outcome of measuring $X$ is one of its eigenvalues $x$---which one, is random.
Its probability distribution (law) depends on $X$ and on $\ket\psi$.
\end{myassumption}
\noindent I write $\meas_\psi(X)$ for the random outcome of measuring 
observable 
$X$ on state $\ket\psi$, and $\law\bigl(\meas_{\psi}(X)\bigr)$ for
its probability distribution, i.e., the collection of probabilities 
$\Pr\bigl\{\meas_\psi(X) \in  B\}$ for all Borel sets $B$ of the real 
line. Deutsch's paper has the more modest aim just to compute the 
mean 
value of this probability law, $\E\bigl(\meas_{\psi}(X)\bigr)$, though 
as I shall argue before, even under his own terms (computing values 
of 
betting games) the whole probability law is of interest.

Throughout the paper I will be playing with three main assumptions,
though sometimes in stronger and sometimes in weaker forms.
Here are the three, in their strongest versions:

\begin{myassumption}\label{a:degeneracy}\textit{Degeneracy in 
eigenstates.}
\begin{equation}
    \Pr\bigl\{\meas_{\ket{X\!=\!x}}(X)~=~x\bigr\}~=~1.
\end{equation}
\end{myassumption}
\noindent
In an eigenstate of an observable, 
the corresponding eigenvalue is the certain outcome of measurement.

\begin{myassumption}\label{a:functional}\textit{Functional invariance.}
\begin{equation}
\Pr\bigl\{f(\meas_\psi(X))~=~y\bigr\}
~=~\Pr\bigl\{\meas_\psi(f(X))~=~y\bigr\}.
\end{equation}
\end{myassumption}
\noindent
Measuring a function $f$ of an observable is operationally 
indistinguishable from measuring the observable, and then taking the 
same function of the outcome. Parenthetically remark that this 
indistinguishability is only as far as the outcome is concerned;  
as far as the new state of the quantum system is concerned 
there will be a difference, if the function is many-to-one. Parts of 
Deutsch's proof only need this assumption for one-to-one functions.
In fact he only explicitly used this assumption for the affine  
functions $f(x)=ax+b$, but implicitly other functions, including 
many-to-one functions, are involved too.

\begin{myassumption}\label{a:unitary}\textit{Unitary invariance.}
\begin{equation}
\Pr\bigl\{\meas_{U\psi}(X)~=~x\bigr\}
~=~\Pr\bigl\{\meas_\psi(U^{*}XU)~=~x\bigr\}.
\end{equation}
\end{myassumption}
\noindent
We will see that, at first instance, we only require this assumption to hold
for a special class of 
unitary operations $U$, namely those which permute eigenstates of $X$.
There is then a one-to-one correspondence $u$ on the eigenvalues of 
$X$ with inverse $u^{*}$ such that $UXU^{*}=u(X)$,
$U^{*}XU=u^{*}(X)$, and $U\ket {X\!=\!x}=\ket{X\!=\!u(x)}$. 
In the special case that $\psi$ is
an eigenstate  $\ket {X\!=\!x}$, Assumption \ref{a:unitary}
follows from Assumption \ref{a:degeneracy}
(degeneracy-in-eigenstates). Later we also need 
Assumption \ref{a:unitary} for unitary operations, diagonal in the basis 
corresponding to $X$.

Since in the above assumptions, $x$ and $y$ are arbitrary, one could 
restate the three main assumptions as:
\begin{align*}
\law\bigl(\meas_{\ket{X\!=\!x}}(X)\bigr)~&=~\law\bigl(x\bigr),\tag{1${}'$}\\
\law\bigl(f(\meas_\psi(X))\bigr)~&=~\law\bigl(\meas_\psi(f(X))\bigr),\tag{2${}'$}\\
\law\bigl(\meas_{U\psi}(X)\bigr)~&=~\law\bigl(\meas_\psi(U^{*}XU)\bigr),\tag{3${}'$}
\end{align*}
where $\law$ denotes the probability law of the random variable in 
question, 
so that in particular $\law(x)$ denotes the probability distribution 
degenerate at the point $x$. An apparently weaker still 
set of assumptions would only restrict the mean values of the 
distributions in Assumptions \ref{a:functional} and \ref{a:unitary}:
\begin{align*}
\E\bigl(f(\meas_\psi(X))\bigr)~&=~\E\bigl(\meas_\psi(f(X))\bigr),\tag{2${}''$}\\
\E\bigl(\meas_{U\psi}(X)\bigr)~&=~\E\bigl(\meas_\psi(U^{*}XU)\bigr).\tag{3${}''$}
\end{align*}
As mentioned above, one can weaken the assumptions by
restricting the class of functions $f$ or unitaries $U$ for which
the relevant equalities are supposed to hold.

\section{The first part of the proof}\label{s:proof}

I return to a discussion of the assumptions after an outline of the 
proof of my main result: 
\begin{equation}
\Pr\bigl\{\meas_\psi(X)~=~x\bigr\}~=~\bigl | \braket{\psi} {X\!=\!x} 
\bigr |^{2}.
\end{equation}
I will make use of Assumptions 1--3 in their original form,
postponing discussion of how one might reach the same 
conclusion from weaker versions of the assumptions.
In this section, following Deutsch, 
I only prove the result in the special case (a)
\begin{equation}
\ket\psi={\textstyle{\frac1{\sqrt 2}}}\bigl(\,\ket {X\!=\!x_{1}}+\ket 
{X\!=\!x_{2}}\,\bigr),
\end{equation}
for which I am going to obtain the probabilities $1/2$ for $x=x_{1}$ 
and 
$x=x_{2}$, and zero for all other possibilities. After this, Deutsch
attempts to generalize, first (b) to equal weight superpositions of a 
binary 
power of eigenstates of $X$, next (c) to an arbitrary number, then (d)
to
dyadic rational superpositions, next (e) to arbitrary real 
superpositions, and finally (f) to arbitrary superpositions. 
The proofs he gives of these steps are similarly 
incomplete. I will complete the proof by an alternative and rather short route
in the next section,
but return to Deutsch's completion in the section after that.

Suppose $u$, a one-to-one correspondence on the eigenvalues of $X$,
maps $x_{1}$ to $x_{2}$ and \textit{vice-versa}, and, after we have labelled 
the other eigenvales as $x'_{n}$, $n \in \mathbb Z$, 
maps $x'_{n}$ to $x'_{n+1}$. Let $U$ denote the unitary
which performs the same permutation of the eigenvectors.
Let $u^*$ denote the inverse of $u$.
Exploiting the relationship between $u$ and $U$, 
and their relationship to $X$ and $\psi$, as well as 
our  other assumptions,
we find,
\begin{align}
\Pr\bigl\{\meas_\psi(X)~=~x_{1}\bigr\}~
    &=~\Pr\bigl\{\meas_{U\psi}(X)~=~x_{1}\bigr\}\notag\\
    &=~\Pr\bigl\{\meas_\psi(U^{*}XU)~=~x_{1}\bigr\}\notag\\
    &=~\Pr\bigl\{\meas_\psi(u(X))~=~x_{1}\bigr\}\notag\\
    &=~\Pr\bigl\{u(\meas_\psi(X))~=~x_{1}\bigr\}\notag\\
    &=~\Pr\bigl\{\meas_\psi(X)~=~u^{*}(x_{1})\bigr\}\notag\\
    &=~\Pr\bigl\{\meas_\psi(X)~=~x_{2}\bigr\}.
\end{align}
Replacing $x_{1}$ by an eigenvalue $x'_n$, i.e., any other than
$x_{1}$ or $x_{2}$, 
and running through the same derivation, we see that all other 
eigenvalues have equal 
probabilities. Since there are an infinite number of them,  and
since  according to our 
background assumption the outcome of measuring $X$ lies in its 
spectrum, we have obtained the required result:
the probabilities of $x_1$ and $x_2$ must both equal $1/2$, all
the other eigenvalues $x'_n$ must get zero probability.

We used Assumptions \ref{a:functional} and \ref{a:unitary}
(functional and unitary invariance), not Assumption 
\ref{a:degeneracy} (degeneracy in an eigenstate). 
However, this assumption is needed to deal with 
the case of  \dots\ an eigenstate. The proof method allows us 
to deal with an equal weight superposition of any positive
finite number of eigenstates of $X$. We only used functional 
invariance for one-to-one functions.

Deutsch was only interested in mean values of the probability 
distributions of outcomes, since the fair value of the game: measure 
$X$ on $\ket\psi$ and receive the value of the outcome in euro's 
(\Euro), is precisely \Euro\ $\E\bigl(\meas_\psi(X)\bigr)$. (Here we 
are 
assuming that the utility of having some number of euro's is equal to 
that 
number. The reader may replace euro's by dollars, camels, or whatever 
else 
he or she prefers). In a moment I will also add a new game to the 
discussion: 
measure $X$ on $\ket\psi$ and receive \Euro\ $1$ if the outcome 
$x_{0}$ is 
found. The value of this game should be \Euro\ 
$|\braket{x_{0}}{\psi}|^{2}$.

Let us assume that the spectrum of $X$ consists of 
\emph{all} the integers (negative and non-negative). 
Then for given $x_{1}$ and $x_{2}$ there is an affine 
map $u(x)=ax+b=x_{1}+x_{2}-x$ which defines a unitary transformation 
$U$ as above. For these $U$, $X$ and the same $\psi$ as before we 
rewrite the 
argument before as
\begin{align}
\E\bigl(\meas_\psi(X)\bigr)~
    &=~\E\bigl(\meas_{U\psi}(X)\bigr)\notag\\
    &=~\E\bigl(\meas_\psi(U^{*}XU)\bigr)\notag\\
    &=~\E\bigl(\meas_\psi(u(X))\bigr)\notag\\
    &=~\E\bigl(u(\meas_\psi(X))\bigr)\notag\\
    &=~x_{1}+x_{2}- \E\bigl(\meas_\psi(X)\bigr)
\end{align}
yielding the required equality,
\begin{equation}
\E\bigl(\meas_\psi(X)\bigr)~=~{\textstyle{\frac12}}(x_{1}+x_{2}).
\end{equation}

Deutsch's proof was a cryptic version of the argument I have 
just given, except that he did not mention the unitary invariance 
assumption. He writes $v$ for value, instead of $\E$.
In my opinion, without the extra (unitary invariance) assumption,
his proof fails. The degeneracy Assumption \ref{a:degeneracy} 
is not used at this stage. However one
may note that Assumption \ref{a:degeneracy} (degeneracy) implies that 
Assumption \ref{a:unitary} (unitary invariance) holds when the state
$\ket\psi$ is an eigenstate of the observable $X$. One could therefore
consider Assumption \ref{a:unitary} as a natural interpolation from 
Assumption \ref{a:degeneracy}. I return to this later.

As has been shown by de Finetti and by Savage, a rational decision 
maker who must make choices when outcomes are `indeterminate' 
(I  must 
avoid all terminology suggestive of probability theory, since the words
`random', `probability' and so on, are not allowed to be in our 
vocabulary) behaves as if he (or she) has a prior probability 
distribution and indeed updates it according to Bayes' law when new 
information (outcomes) becomes available. Thus it seems to me that 
whether one starts with utilities and assumes rationality, or with 
probability and the frequency interpretation, is very much a matter 
of taste. In my opinion the latter is closer to physical experience 
and indeed we know that casinos and insurance companies make good 
money from the frequency interpretation of chance.

I consider the many repetitions in the frequency interpretation to be 
no more and no less than a thought experiment. 
When one claims that the probability of some event is some number, 
one is asserting that the situation in question is indistinguishable 
from a certain roulette game or lottery. This allows me also to talk 
about probabilities of outcomes of once-off experiments. For 
instance, a certain physical experiment might have some chance of 
producing a black hole which would swallow the whole universe. The 
probability that this would indeed happen, if the devilish experiment 
were actually carried out, would be computed by doing real physics in 
which one would imaginarily set the chain of events into motion, many 
many times, in which uncontrolled initial conditions would vary in 
all kinds 
of ways from repetition to repetition. How they would vary, and what
possibilities could be considered equally likely, should be a matter 
of scientific discussion. This may appear circular reasoning  or an
infinite regress or just plain subjectivism, but this does not bother 
me: 
it works, and it is not subjective, since we may rationally discuss 
the probability modelling. When I use the mathematical model of 
probability, I am only 
claiming an analogy with something familiar, like a casino, lottery, 
or coin 
toss. I think that it is the same in the rest of physics, when we 
talk about mass, electric charge, or magnetic field: we might think 
or we might hope that we are talking about real things in the real 
world 
but we can only be certain that we are talking about ingredients of 
mathematical models which are anchored to the real world by analogies 
with familiar down to earth daily experience. My frequentistic 
position 
is perhaps better labelled ``Laplacian counterfactual 
frequentism'' and though one might collapse this 
label to ``subjectivism'', I believe it is as instrumentalistic or 
as operationalistic as anything else in physics.

\section{Completing the proof}\label{s:rest}

More can be got out of the functional invariance assumption, 
by considering other functions $f$, and most crucially, certain 
many-to-one functions. In my opinion we must do this anyway,
in order to complete the proof on the lines indicated 
by Deutsch (see next section). 
It is an open question, whether we can do without. 

With the choice $f=\onebb _{\{x\}}$, and writing $[X\!=\!x]$ for the 
projector onto the eigenspace of $X$ corresponding to eigenvalue $x$
(and later also for the eigenspace itself), since $\onebb 
_{\{x\}}(X)~=~ [X\!=\!x]$, we read off:
\begin{equation}
\Pr\bigl\{\meas_\psi(X)~=~x\bigr\}~=~\Pr\bigl\{\meas_\psi 
([X\!=\!x])~=~1\bigr\}.
\end{equation}
Indeed, if we only assume the mean value form of the functional 
invariance assumption, we can read off the same conclusion, since 
the random variables $\onebb _{\{x\}}(\meas_\psi(X))$ and
$\meas_\psi([X\!=\!x])$ are both zero-one valued.

Till this point we had dealt with nondegenerate observables and equal 
weight 
superpositions of eigenstates. Now we can add to this, also 
degenerate observables (since these can always be written as 
functions of nondegenerate observables). Moreover, even if we start 
with the 
assumptions in their weaker mean value form, we can still obtain the 
stronger conclusion about the whole probability law of the outcome.

In fact, with brute force we arrive now very quickly at the most 
general 
result (it remains, namely, to consider arbitrary states). 
From functional invariance 
(whether in terms of probability laws or whether in terms of their 
mean values) 
we have shown that a probability can be assigned to each closed 
subspace 
of our Hilbert space, countably additive over orthogonal subspaces, 
and equal 
to $1$ on the whole space. Now we can invoke Gleason's theorem to 
conclude that 
the probability of any subspace is of the form $\mathrm{tr}\{\rho 
A\}$ for some density matrix $\rho$. It remains to show that 
$\rho=\ket\psi\bra\psi$ but this follows from our first axiom that 
measuring an observable on an eigenstate yields with certainty the 
corresponding eigenvalue: consider the observable 
$X=\ket\psi\bra\psi$ 
itself, and subspace $A=[\psi]$ (the one-dimensional subspace 
generated by $\ket\psi$)! 

Deutsch's extension of his results to the most general case
(see next section) is very hard to follow. 
He repeatedly invokes substitutability, 
whereby 
an outcome of one game may be replaced by a new game of the same 
value.
He does not say \emph{which} substitutions are being made. However he 
is clearly thinking of substitutions, leading to composite games with 
composite quantum systems, product states, and observables on each 
subsystem.  During these constructions and substitutions, the 
observables being measured and the states on which they are being 
measured, keep changing, while the Spartan notation $v(x)$ in which 
the 
symbol $x$ refers to an observable, an eigenvalue, and an eigenstate 
simultaneously, begs confusion. The mere construction of product 
systems implies that more is being assumed above the structure so far
(so far we only spoke of observables and states on one fixed quantum 
system). As I will indicate below, it appears that the extra 
assumption of unitary invariance and the strengthened functional 
invariance 
assumption involving many-to-one functions as well as one-to-one 
functions, together with a natural assumption about measuring 
separate 
observables on a product system in a product state, enable one to 
fill 
the gaps. If the repair job is not too difficult, one finishes  
with a 
relatively easy proof of Gleason's theorem, under the supplementary 
condition of unitary invariance. 

The construction of product systems will also help us extend results 
from infinite-dimensional quantum systems to finite dimensional, 
including $2$-dimensional---the case not covered by Gleason.

Functional invariance assumptions on product systems, or more 
generally, for compatible observables, play a 
key role in many foundational discussions of quantum mechanics. 
Recall that 
observables $X$, $Y$ commute (or are compatible with one another) 
if and only if both are functions of a third
$Z$; and the third can be chosen in such a way (with a minimal set of 
eigenspaces) to make the mapping $Z\mapsto (X,Y)$ a one-to-one 
correspondence in the 
sense that we can write $X=f(Z)$, $Y=g(Z)$, $Z=h(X,Y)$ where $h$ is 
the inverse of $(f,g)$. In other words, two (or more) commuting 
observables can be thought of as components of a vector-valued 
observable, or equivalently as defining together one `ordinary' 
observable.
Whether one thinks of them together as a vector or as a scalar observable 
is merely a question of how the eigenspaces are labelled.
One can define joint measurement of compatible observables in several 
equivalent ways. Assuming L\"uders' projection 
postulate for how a state changes on measurement, the sequential 
measurements, in any order, of a collection of compatible 
observables, 
are operationally indistinguishable from one another. 
One may therefore think equally well of `one-shot' 
measurement of $Z$, sequential measurement of $X$ then $Y$, and 
sequential measurement of $Y$ then $X$.

This leads to a further extended functional invariance assumption:
\begin{equation}
\law\bigl(f(\meas_\psi(\vec X))\bigr)~=~\law\bigl(\meas_\psi(f(\vec 
X))\bigr),
\end{equation}
where $\vec X=(X_{1},\dots,X_{k})$ is a vector of mutually compatible 
observables and $f:{\mathbb R}^{k}\to {\mathbb R}^{m}$. Apparently 
weaker 
is the mean value form of this:
\begin{equation}\label{e:unconsc}
\E\bigl(f(\meas_\psi(\vec X))\bigr)~=~\E\bigl(\meas_\psi(f(\vec 
X))\bigr);
\end{equation}
though as I showed above, by playing around with indicator functions,
the two are equivalent. We can recover from the assumption the 
fact that the probability law of a measurement 
of $X$ alone is the same as the first marginal of the joint law of 
the two outcomes of a joint measurement of commuting $X$, $Y$.
As I have argued  in Gill(1996a,b), these 
consequences of the standard
theory form a crucial though often only implicit ingredient 
in many of the famous no-go arguments against hidden variables in the 
literature. Somewhat irreverently I have dubbed (\ref{e:unconsc}) 
`the law of the unconscious quantum physicist'. 

Deutsch's approach is similar to that of some probabilists, in that 
he 
would prefer to make Expectation central, and have Probability a 
consequence (in fact, he would prefer to do without the word 
Probability altogether). This is fine, and indeed many probabilists 
do 
take this approach (Whittle in his textbook on Probability argues 
that one should do the same for quantum probability, too).
Now in our situation we want to start with hypothesizing existence of 
mean values, and by making some structural assumptions about them. 
From 
this we want to derive the form of the mean values. As I have noted 
above, since $\onebb_{\{x\}}(X)$ is a both an observable itself, 
and a function of the observable $X$, it would appear that fixing all 
mean values of (outcomes of measurements of) all observables, fixes 
all 
probability laws of (outcomes of measurements of) all observables. 
The point I want to make, is that this indeed works, provided we have 
the 
functional invariance assumption (for mean values only, if you like, 
but we must have if for a very large class of functions). Do we need 
to consider many-to-one functions? If our assumptions are 
\emph{only} about expectations, I think we do need many-to-one 
functions. However, with modest
distributional input, one need further only consider one-to-one 
functions,
as follows. Suppose we know 
the mean value of $\meas_{\psi}(\exp(it \arctan X))$, and suppose we 
assume 
functional invariance, \emph{in law}, for all one-to-one functions;
in particular, the functions $f(x)=\exp(it \arctan x)$, for each real $t$.
Then we know $\law(\meas_{\psi}X)$. It is possible to avoid 
complex-valued
functions, try for instance $f(x)=s\cos(\frac12(\arctan x+\pi/2))+
t\sin(\frac12(\arctan x+\pi/2))$ for all real $s$ and $t$.

Let me return to the contrast between Deutsch's and Gleason's 
argument.
Deutsch's proof, on completion, seems a little simpler and more 
direct. 
His assumptions are much stronger: he needs unitary 
invariance. His assumptions are more representative of classical 
quantum 
mechanics---unitary evolution \emph{has} to be considered an 
essential 
part of this. In the first stages of his argument, deriving mean 
values for
some rather special observables and rather special states, he 
moreover only needed to consider functional invariance under 
one-to-one transformations. This assumption is close to tautological 
(the 
apparatus for measuring $a+bX$ is not going to be essentially 
different from that for measuring $X$). However, even from the point 
of view of deriving fair values of games, probability laws as well as 
mean values are equally relevant. For instance, what is 
the fair value of the game: measure $X$ and receive \Euro\ $1$ if the 
outcome $x_{0}$ is obtained? The easiest way to deal with this game 
too, 
is to include functional invariance for the indicator functions too, 
and then one need not work any more but simply appeal to Gleason's 
theorem.

\section{Discussion}\label{s:discuss}

Later in this section I will run through Deutsch's steps
to complete his proof. The aim will be to see whether, with weaker versions of
our main assumptions, not strong enough to give us Gleason's assumptions
so easily, we could also arrive at the desired conclusion. (The answer is
that at present, I do not know).
But first I would like to discuss what grounds one could have for the 
functional and unitary invariance assumptions, against the background 
assumptions that measuring an observable yields an eigenvalue, and 
that in an eigenstate, the outcome is certain. 

Functional invariance 
for one-to-one functions seems to me more or less 
definitional. For many-to-one it is much less definitional, 
also less empirical, since there will vary rarely truely exist 
essentially different measurement apparatuses for `doing' $X$ and 
doing $f(X)$. Just occasionally there will be empirical evidence 
\emph{supporting} functional invariance: for instance, when $X$ and 
$Y$ do not 
commute, but for some many-to-one functions, one has $f(X)=g(Y)$, 
there might be empirical (statistical) data supporting it, 
based on the quite different experiments 
for measuring $X$ and for measuring $Y$, and finding the same 
statistics (or mean values) for $f$ of the outcomes of the first 
experiment, $g$ of the outcomes of the second. There is one very 
strong empirical fact supporting the assumption (in its form for 
vector observables): when we simultaneously measure observables on 
separate 
components of a product system (even if in an entangled state)
we have the same marginal statistics, as if only one component was 
being 
measured. Altogether, the nature of this assumption would seem to me 
to be: 
we extend a definitional assumption concerning a smaller class of 
functions 
$f$---the affine functions---to a much larger class, by mathematical 
analogy, 
trusting that the world is so elegantly and mathematically put 
together, 
that the `obvious' sweeping mathematical generalization of an 
indubitable 
fact is usually correct; we are supported in this by some empirical 
(statistical) evidence for some special cases.

Similarly the assumption of unitary invariance seems to be 
largely a leap of faith, since there will be little empirical 
(statistical) evidence to support it. But again, 
one might prefer to think of the leap of faith 
as a natural mathematical generalization. Our first assumption---that 
measuring an observable on an eigenstate produces the 
eigenvalue---tells us
\begin{equation}
\law\bigl(\meas_{U\psi}(X)\bigr)~=~\law\bigl(\meas_\psi(U^{*}XU)\bigr),
\end{equation}
whenever $U$ permutes eigenspaces and $\psi$ is an eigenvector!
Extending this to arbitrary states can be thought of as an 
interpolation, in harmony with ideas of wave-particle duality.
It seems to me that wave-particle duality---the very heart of quantum 
physics---essentially forces probability on us, since it is the only 
way to get a smooth interpolation between the distinct discrete 
behaviours at different eigenstates of an observable. We just have to 
live with smoothness at the statistical level, instead of at the 
(counterfactual) level of individual outcomes.

I would now like to discuss the remaining steps of 
Deutsch's proof. As we saw, functional invariance in its strongest 
form implies the conditions of Gleason's theorem, which makes all 
further conditions and further work superfluous. Now the reason 
functional 
invariance is so powerful, is that we assumed it to hold 
for \emph{all} functions $f$, in particular, many-to-one functions. 
In 
the spirit of the first part of Deutsch's proof it would make sense 
to demand it only for 
\emph{one-to-one} functions. It seems to me a reasonable conjecture 
that Deutsch's theorem is true under the three assumptions: 
functional invariance for one-to-one functions, unitary invariance, 
and the degeneracy assumption.

As was stated earlier, after (a) the two-eigenstate equal weight 
superposition, Deutsch extends this (b) to binary powers, (c) to 
arbitrary whole numbers of equal weight superpositions, (d) to 
rational superpositions, (e) to real and finally (f) to arbitrary.
As we saw, steps (b) and (c) can also be dealt with by  
his own method for the two-eigenstate case. Deutsch's argument for (d) 
involves completely new ingredients and assumptions. He supposes that
an auxiliary quantum system can be brought into interaction with the 
system under study, thus yielding a product space and a product state.
The observable of interest $X$ is identified with $X\otimes\mathbf 1$,
and this is considered as one of a pair $(X\otimes\mathbf 1,\mathbf 
1\otimes Y)$ where the observable $Y$ is cleverly chosen, so that 
in the product system, and with this product observable, we are back 
in an equal weight superposition of eigenstates. 
He then makes the assumption: measuring $X$ on the 
original system is the same as measuring $(X,Y)$ on the product sytem 
and discarding the outcome of $Y$. Uncontroversial though this may 
be, 
we are greatly expanding on the background assumptions. Moreover we 
are actually assuming functional invariance for a many-to-one 
function: namely, the function which delivers the $x$-component of a 
pair $(x,y)$. By the way, Deutsch's proofs of steps (b) and (c) 
similarly 
involve such constructions. Step (e) is an approximation argument 
which can presumably be made rigorous, though perhaps differently to 
how Deutsch does it. Step (f) as 
presented by Deutsch involves yet another new assumption: 
measuring an 
observable can be represented as a unitary transformation on a 
suitable product system, so that after a new unitary transformation 
mapping $\ket x$ to $e^{-i\phi}\ket x$ one can remove complex phases 
from a superposition of eigenstates. This argument seems to be unnecessarily 
complicated. Our unitary invariance assumption 
together with the unitary transformation just described, takes care 
of 
extending results from real to complex superpositions.

The work of Deutsch has been strongly criticised by Finkelstein 
(1999) and by  Barnum \textit{et al}\ (1999). They 
also point out that the first step of Deutsch's proof is incorrect, 
however, do not recognise that it can be repaired by a supplementary, 
natural, condition. They also point out that Gleason's theorem does 
the same job as Deutsch purports to do, but do not see the very close 
connection between Gleason's and Deutsch's assumptions. They point 
out also that the later steps of Deutsch's proof depend on various 
appeals to the substitutability principle, without stating which 
games 
were to be substituted for which. I must admit that 
it took me a long email correspondence with David Deutsch, before I 
was able for myself to fill in all the gaps. Finally they also point 
out that the work of de Finetti and Savage implies that  
rational behaviour under uncertainty implies behaviour as if 
probability is there. It is therefore just a question of taste 
whether or not one adds a probability interpretation to the `values 
of games' derived by Deutsch.

My conclusion is that Deutsch's proof as it stands is valid, 
though the author is implicitly using unitary as well as functional 
invariance. All his assumptions together imply the assumptions of 
Gleason's theorem, and much more. Consequently the proof as given 
does 
not have a great deal of mathematical interest. However the fact that 
distributional conclusions could already be drawn for \emph{some} 
states and \emph{some} observables, at a point at which only 
functional invariance for one-to-one functions had been used, and in 
my opinion, with a most elegant argument, justifies the 
conjecture I have already mentioned: 

\begin{myconjecture}
Deutsch's theorem is true under 
the three assumptions: functional invariance for one-to-one 
functions, 
unitary invariance, and the degeneracy assumption.
\end{myconjecture}

Unitary invariance alone tells us that the law of the outcome of a 
measurement of $X$ only depends on the absolute innerproducts 
$|\braket x \psi |$.
So the task is to determine the form of the dependence.

\section*{Acknowledgments}

I am grateful for the warm hospitality and support  of the Quantum Probability 
group at the department of mathematics of the University of Greifswald, 
Germany, during my sabbatical there, Spring 2002. 
My research there was supported by European Commission grant 
HPRN-CT-2002-00279, RTN QP-Applications.
This research has also been supported by project RESQ (IST-2001-37559) 
of the IST-FET programme of the European Commission.


\begin{thebibliography}{0}

\raggedright
    
\bibitem[\protect\citeauthoryear{Barnum, Caves, Finkelstein, Fuchs 
and Schack}{Barnum \textit{et al}.}{2000}] {barnumetal00}
H. Barnum, C. M. 
Caves, J. Finkelstein, 
C. A. Fuchs and R. Schack (2000),
Quantum Probability from Decision Theory?
\textit{Proc.\ Roy.\ Soc.\ Lond.\  Ser. A} {\bf 456}, 1175--1182.

    
\bibitem[\protect\citeauthoryear{Deutsch}{Deutsch}{1999}]{deutsch99} 
D.~Deutsch (1999), Quantum Theory 
of Probability and Decisions, \textit{Proc.\ Roy.\ Soc.\ Lond.\ Ser.\ A} 
\textbf{455}, 3129--3137.

\bibitem[\protect\citeauthoryear{Gill}{Gill}{1996a}]{gilla}
R.D. Gill (1996a), Discrete Quantum Systems,
\texttt{www.math.uu.nl/people/gill/Preprints/chapter2.pdf}.

\bibitem[\protect\citeauthoryear{Gill}{Gill}{1996b}]{gillb}
R.D. Gill (1996b), Hidden Variables and Locality,
\texttt{www.math.uu.nl/people/gill/Preprints/chapter14.pdf}.

\bibitem[\protect\citeauthoryear{Gleason}{Gleason}{1957}]{gleason}
A. Gleason (1957), Measures on closed subsets of a Hilbert space,
    \textit{J. Math.\ Mech.} {\bf 6}, 885--894.

\bibitem[\protect\citeauthoryear{Saunders}{Saunders}{2002}]{saunders}
S. Saunders (2002), Derivation of the Born Rule from
Operational Assumptions,
\texttt{quant-ph/0211138}.

\bibitem[\protect\citeauthoryear{Wallace}{Wallace}{2002}]{wallace}
D. Wallace (2002), Quantum Probability and Decision Theory, Revisited,
\texttt{quant-ph/0211104}.


\bibitem[\protect\citeauthoryear{Wallace}{Wallace}{2003}]{wallace2}
D. Wallace (2003a), Everettian Rationality: defending Deutsch's approach 
to probability in  the Everett interpretation,
\texttt{quant-ph/0303050}.
To appear in \textit{Studies in the History and Philosophy  of Modern Physics}, 
under the title ``Quantum Probability and Decision Theory, Revisited''.


\bibitem[\protect\citeauthoryear{Wallace}{Wallace}{2003}]{wallace3}
D. Wallace (2003b),Quantum Probability from Subjective Likelihood: 
improving on Deutsch's  proof of the probability rule,
\texttt{quant-ph/0312157}.



\end{thebibliography}
\end{document}